# Ambient Pressure-Driven Actuators and Shape-Morphing Metamaterials

Benjamin A. Holdstock*, Robert Trowbridge, Julian Londono Monsalve, Alexander W. Powell

Centre for Metamaterials Research and Innovation, Physics Building, Streatham Campus, University of Exeter, Stocker Road, Exeter, United Kingdom, EX4 4QL

Keywords: Shape-morphing, biomimetics, expansivity, actuator, ambient, pressure, negative compressibility, bi-layer, metamaterial

Abstract graphic:

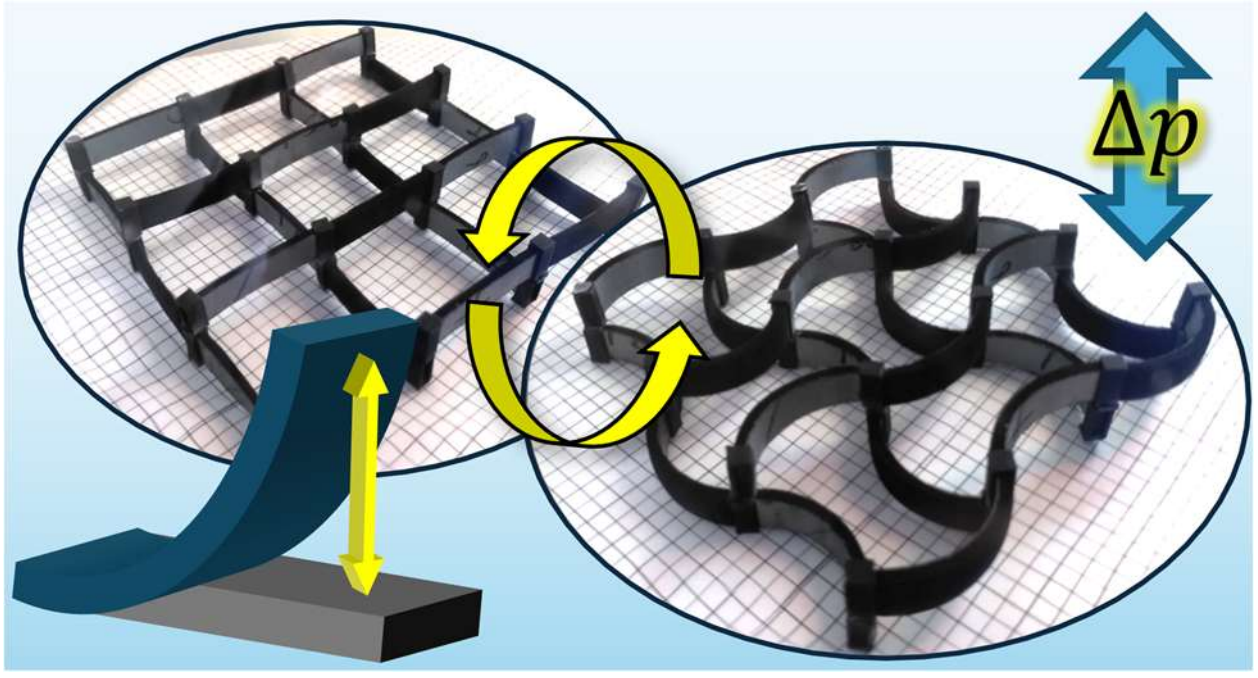

Abstract:

Actuators which change shape in response to ambient pressure have great potential for use in extreme environments, from space to the deep ocean. Bi-layer actuators have so far not been explored to this end, despite their lightness and mechanical simplicity (there is no requirement for a separate power source or pump, for example) making them ideal for applications where transport and maintenance are difficult. In this paper, foam-based bilayer actuators powered by ambient pressure change have been constructed, modelled mathematically, characterised in terms of their actuating force, and built into deployable devices and metamaterial structures to showcase properties such as negative pressure expansivity. Actuators such as these are ideally suited for a range of applications, from deploying solar panels and sensors in space and extra-terrestrial atmospheres, to under-sea expandable structures and sensing.

## 1 Background

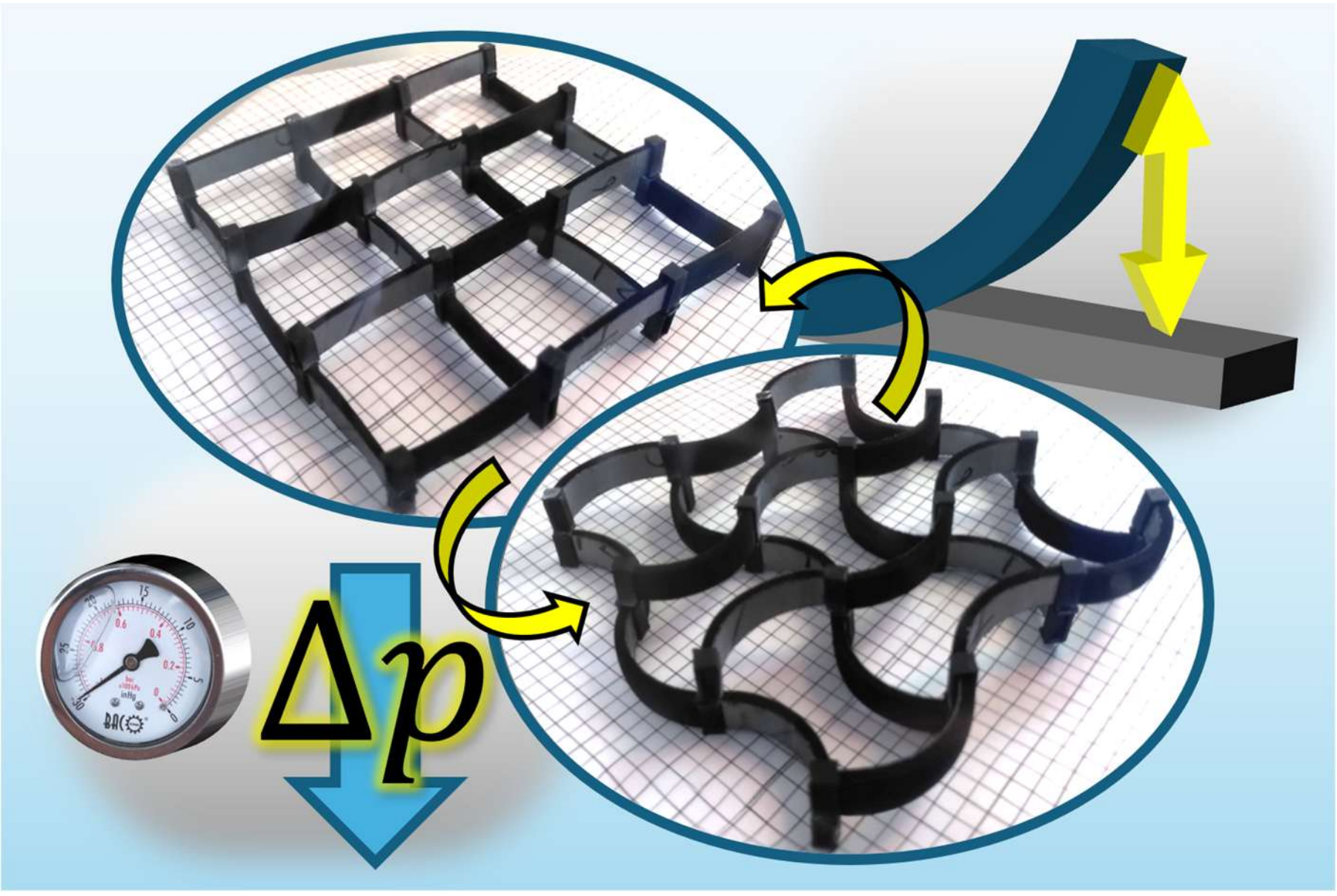

**Figure 1.** A visualisation of the actuator bending process as the ambient pressure is varied, both in isolation and as part of a metamaterial mesh.

Mechanisms which move require actuators. These exist in many forms; electrical actuators are some of the most widespread, which, while adaptable and highly tuneable, can be heavy and complex in comparison to the devices they are required to move. This is due to the requirements for both control electronics and a power source. While less electrically complex, hydraulic and pneumatic actuators [1] both also require driving sources and air- (or water-) tight connections. Nature offers a simpler solution - instead of a pre-programmed electrical impulse, changes in environmental conditions can be used to trigger shape change (and thus motion). This mechanism is widely seen in the natural world, in a variety of both plant [2-5] and animal [6,7] anatomies.

In order to demonstrate an advantage over conventional electrical or pneumatic/hydraulic actuators, any replacement should be as simple and light as possible, with no requirement for an external driving source. Shape-memory polymers (SMPs) have been employed for this purpose, which can utilise a variety of environmental changes as triggers including heat and light [8-12]. However, SMPs are not easily reversible, since they operate by returning to their original state following an induced deformation – they must therefore be reset before every actuation. Enter the shape-morphing bi-layer: two strips of different materials, bonded together along one face, each with a different response to ambient conditions. The most common, and best understood, form of these is bi-metal thermostats, whose bending behaviour has been well-understood for a hundred years [13]. Since then, polymer bi-layers have been trialled in a variety of settings, from direct replication of biological systems, for example through bending in response to humidity change [14]. Bi-layer actuators have also been constructed which respond to light [15], heat [16], pH [17], and changes in the concentration of specific chemicals in the environment [18].

A previously unexplored application of environmentally driven shape-morphing is in extreme pressure environments, such as space, the deep ocean, and extraterrestrial atmospheres. Here, variation in local pressure can be used to drive shape change, without the need for large, heavy motors or pumps – our preliminary work illustrates their considerable potential [19]. This makes these actuators ideally suited for deploying various pieces of equipment, such as solar panels, solar shields, and antennae, as well as deployable structures and other scientific instrumentation.

In this paper, ambient-pressure actuators, constructed from bi-layers formed of closed-cell rubber foams and flexible polymers, have been modelled and developed for the first time. Equations were derived, relating the ambient pressure change to the length contraction and actuator bending radius. The bending characteristics and lifting force were then experimentally investigated, and the observed behaviours used to inform the creation of a 2D, pressure responsive metamaterial (whose pressure expansion properties can be determined by both the actuators and the metamaterial design – see Figure 1), and a deployable shield device.

## 2 Materials and bending

Ambient-pressure driven bilayer actuators consist of two materials with different ambient pressure expansivities, fixed together along one face. The difference in expansivity determines the level of shape change (and thus actuation). Materials were selected to maximise this difference: a highly expansive layer, formed of flexible, closed cell foam, and a second flexible layer which expands negligibly.

Following the testing of several commercial foams (via expansivity measurements obtained by decreasing the pressure in a vacuum chamber), ethylene propylene diene monomer (henceforth EPDM) rubber foam was selected for use. Experiments were performed in a BacoEng 1.5 Gal aluminium vacuum chamber, connected to the supplied pump. Data were recorded using a Fairphone 6 mobile phone camera pointed through the glass chamber lid, with deflection measurements and radii extracted using ImageJ software [20] (calibrated via squared-paper backing applied to the rear wall of the chamber). The ambient pressure expansivity (fractional length change per kPa) of EPDM, $\varepsilon$, was measured to be $-0.84 \pm 0.04$ MPa$^{-1}$, the highest of all tested, and it was verified that this value was constant across the pressures used in the following experiments ($\varepsilon$ is negative, as the material expands when ambient pressure, $p$ decreases). $\Delta p$, equal to the difference between the measured $p$ and atmospheric pressure $p_0$, was varied within the range $0 > \Delta p > -98$ kPa). In this paper, ε refers to the linear pressure expansivity, defined via ${}^{\mathrm{d}l}/_{\mathrm{d}p} = \varepsilon l$ where $l$ refers to the strip length in the contracted state ($\Delta p = 0$), and $\Delta l$ refers to the change in length between this and the expanded state ($\Delta p < 0$).

To create actuators, a second material was required. To maximise the difference in expansivity between this and the EPDM, a non-expansive film was selected, formed from industrial tape. This was applied to sheets of EPDM in bulk, and individual actuators of various dimensions were then cut using a laser cutter.

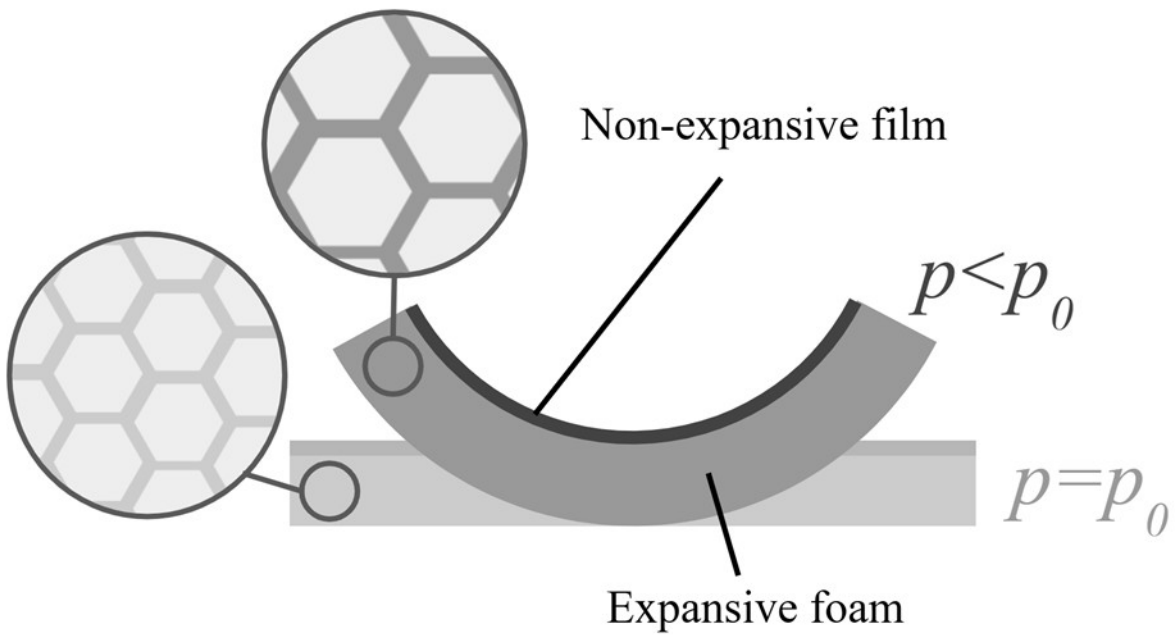

**Figure 2.** The shape-change of a bi-layer actuator, as ambient pressure $p$ decreases from atmospheric pressure ($p_0$). The dimensions of the film (thin, dark) layer are unchanged, while air pockets within the (thick, light) foam layer, and hence the foam as a whole, expand. As the layers are joined along one face, bending occurs.

When $p$ decreases (or increases) from atmospheric pressure $p_0$, the foam component of these actuators expands (or contracts) in all directions. Experiments here were performed by decreasing $p$ via a vacuum chamber, so expansion was observed. The non-expansive film fixes the top surface firmly in place, while allowing the bottom surface to expand freely. It is this which causes the whole bi-layer structure to bend. Where the length-to-width ratio is large (tests showed that 3:2 was sufficient), actuators will experience negligible transverse bending, instead curving longitudinally along a single line to a specific, pressure-driven radius (Figure 2).

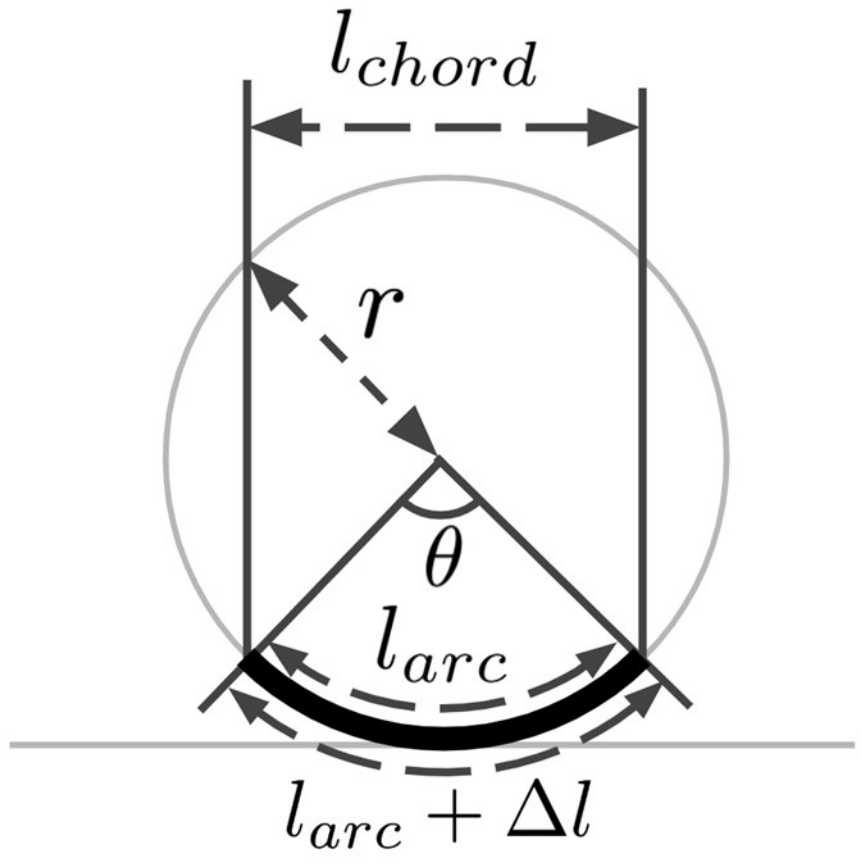

**Figure 3.** Geometric quantities of interest, for an actuator in the curved state (shown in black). As the film does not change shape, $l_{arc}$ (the length of the actuator inner face) is equal to the original length of the actuator before bending. Due to the pressure change, however, the length of the outer face has increased by $\Delta l$. $r$ is the radius of the circle formed by extending $l_{arc}$, with θ corresponding to the angle which $l_{arc}$ subtends. Finally, $l_{chord}$ is the length of the chord across $l_{arc}$.

In the flat state ($\Delta p = 0$), both the top (film) and bottom (foam) faces have the same length, equal to $l$. In the curved state ($\Delta p < 0$), the inner edge remains the same, now redefined as $l = l_{arc}$ for geometric clarity. However, the length of the outer edge increases, to $l_{arc} + \Delta l$. Taking the inner $l_{arc}$ as an arc of a circle, radius $r$, a chord (length $l_{chord}$) may also be defined. θ is defined as the angle subtended by $l_{arc}$ and $l_{chord}$ (Figure 3).

The start point for developing a model was a formulation from Liu et al, designed for non-porous polymer actuators [3]. While this paper focuses on shape-morphing triggered by changes in electric potential, these actuators bend by up to $120\,^{\circ}$ – far closer to the angles seen in this investigation.

An actuator in its bent state may be defined geometrically as follows, where $t$ is the total thickness of the bi-layer (both layers):

$$l_{arc} + \Delta l = (r + t)\theta \qquad (1)$$

$$l_{arc} = r\theta \qquad (2)$$

Two alterations have been made from the original form. Liu et al included an additional term in Equation 2 to represent the length change of the inner layer (the equivalent of $\Delta l$ in Equation 1). However, as in this paper this (film) layer is non-expansive, this term is zero. For the same reason, Equation 1 has also been modified, $\frac{t}{2}$ (representing the position of the interface between the two expansive materials) being replaced by $t$; $\Delta l$ was therefore measured around the outermost edge of the actuator, rather than halfway through.

θ, which is difficult to measure directly, can be eliminated by combining equations 1 and 2:

$$l_{arc} + \Delta l = (r + t)\frac{l_{arc}}{r}, \qquad (3)$$

which simplifies to

$$\Delta l = \frac{l_{arc}t}{r}. \qquad (4)$$

Reintroducing ε (defined previously via $\frac{\mathrm{d}l}{\mathrm{d}p} = \varepsilon l_{arc}$), Equation 4 may be rewritten as

$$r = \frac{t}{\epsilon\Delta p}, \qquad (6)$$

a relationship between $r$ and $\Delta p$ which can be tested experimentally. For the actuators in question, $t = 2.5$ mm, and $r = 31 \pm 2$ mm at $\Delta p = -98$ kPa.

To perform this experiment, actuators of length $50 < l < 100$ mm (in 10 mm increments) and width 25 mm were placed into a vacuum chamber, in front of squared-paper backing. These dimensions were governed by the size of the vacuum chamber; $l$ was maximised within this range, so that shape changes were as pronounced as possible (and hence easier to measure). $w$ was then selected to ensure that the ratio $\frac{l}{w}$ exceeded 1.5 to avoid curvature along the other axis. The pressure in the chamber was gradually reduced, with photographs taken at 10 kPa intervals. Upper- and lower-bound circles were fitted to each photograph; these were taken as the limits of uncertainty for each measurement. The radius of each circle was measured using ImageJ pixel-counting software [20], and calibrated using the squared-paper backing.

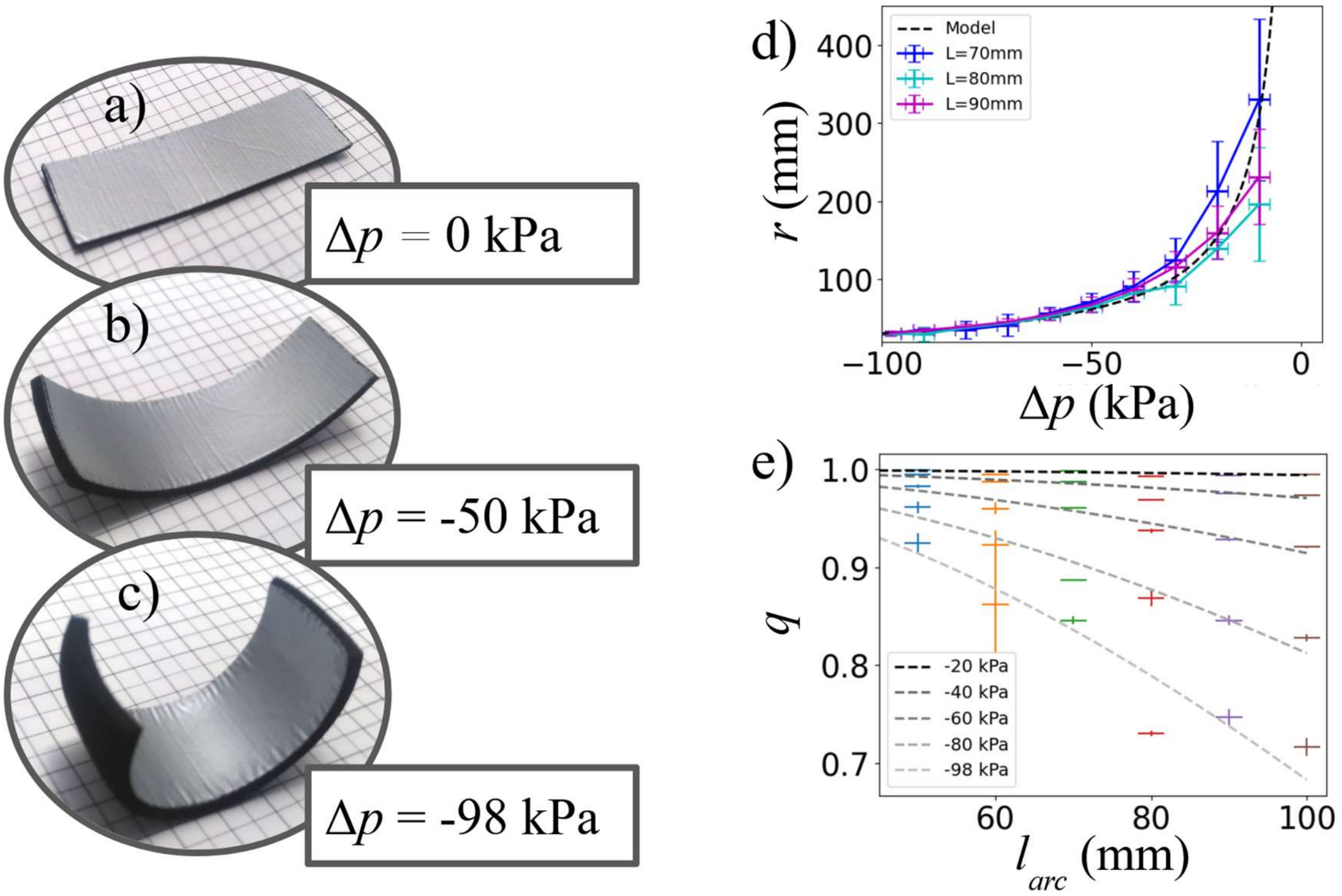

**Figure 4.** An actuator (length $l_{arc}$) at three different pressures $\Delta p$ (a-c), showing increased curvature, and thus a decrease in bending radius. The pressure-radius relation (d) contraction factor $q$ as a function of $l_{arc}$ (e) are also illustrated – for the latter, dotted lines show the model, and coloured plusses show different lengths.

The results of these measurements are shown in Figure 4. No length-dependence was observed in the data (as predicted by Equation 6), and for all points the model lay within the calculated uncertainty range. This was taken as the standard deviation of upper and lower bounds on the radius of circles which could reasonably be fitted to the curved arcs of the actuators. The points themselves were taken as the means of the midpoints between the upper and lower bounds, across four repeats. There is greater uncertainty when fitting circles to small deviations from $\Delta p = 0$, and so a larger experimental error was observed in this case.

For many applications of this material, a more useful measure than the radius of curvature is the fraction by which the spatial distance between the ends of an actuator reduces during bending, the linear contraction factor *q*. This was defined as:

$$q = \frac{l_{chord}}{l_{arc}}. \quad (7)$$

As quantities $l_{chord}$ and $l_{arc}$ are geometrically related by

$$l_{chord} = 2r\,sin\left(\frac{\theta}{2}\right) = 2r\,sin\left(\frac{l_{arc}}{2r}\right), \quad (8)$$

*q* may also be written as

$$q = \frac{2r}{l_{arc}} sin\left(\frac{l_{arc}}{2r}\right), \quad (9)$$

which is a sinc function. Physically, this means that as $l$ increases from 0, $q$ will decrease from 1 until it reaches a minimum – to reach beyond this however, the actuator must pass through $q = 0$, the point at which both ends of the actuator occupy the same point in space. This sets a limit on the validity of the model, $r \geq \frac{l_{arc}}{2\pi}$ – going beyond this point would require a modification, to account for the fact that the actuator ends would be displaced from the diameter of the circle in order to cross over.

To test this model, *q* was calculated for all data points recorded – Figure 4e shows the results. Again, a close correspondence with the model is observed – there is strong correlation between $q$ and $l$ for a given $\Delta p$. The desired $q$ for a given actuator can therefore be precisely optimised by varying its length $l$, as well as its expansivity.

These models can describe more than simple actuators – they can also be used to describe the behaviour of extended metamaterial structures, designed to morph with ambient pressure. To demonstrate this, several 2D metamaterial mesh structures were constructed, inspired by a pneumatic system built from elastomer rather than foam [21].

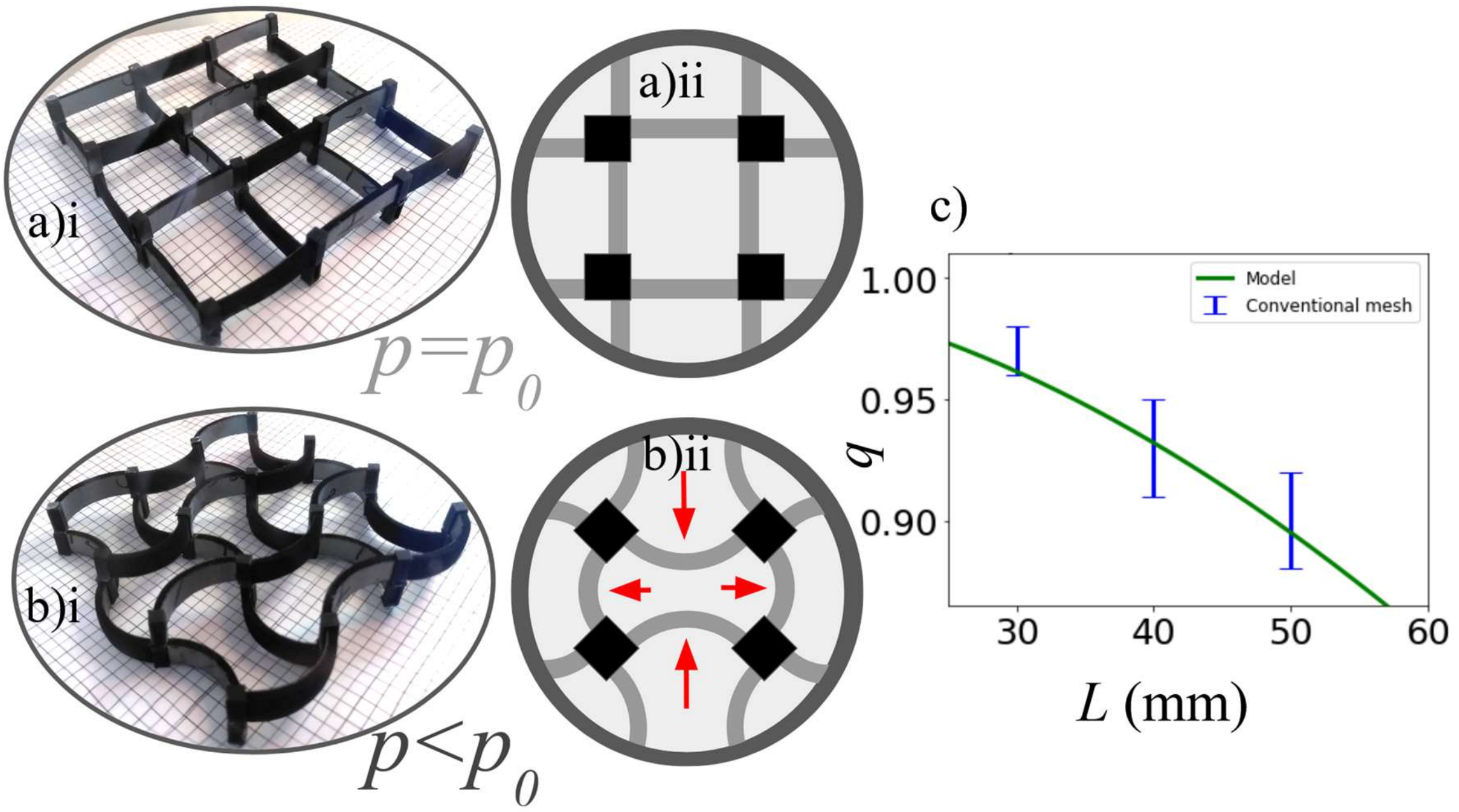

**Figure 5.** A metamaterial mesh, at pressure relative to atmosphere $\Delta p = 0$ (a) and $\Delta p = -98$ kPa (b). Overall shape change is visible (i), and diagrams illustrate bending at the level of individual cells (ii) – red arrows indicate bending direction at actuator center. c) Illustrates that Equation 9 models linear contraction factor $q$ well for an extended structure.

EPDM-film bi-layer actuators were arranged in a grid, with tape affixed to alternating sides, fixed together using 3D-printed ABS nodes at actuator joins (Figure 5). This forms a metamaterial structure similar to another existing SMP design [22], but the use of shape-morphing bi-layers ensures that it can be deformed many times without having to be reset.

Furthermore, as the material itself responds to changes in its environment, no connection to a bulky pump is required for operation.

Individual actuators within the grids were observed to follow the model for $q$ closely - the effect of the structure on the induced radius of curvature $r$ was thus found to be negligible. This allowed the contraction of the meshes as a whole to be predicted accurately, using Equations 6 and 9 without modification.

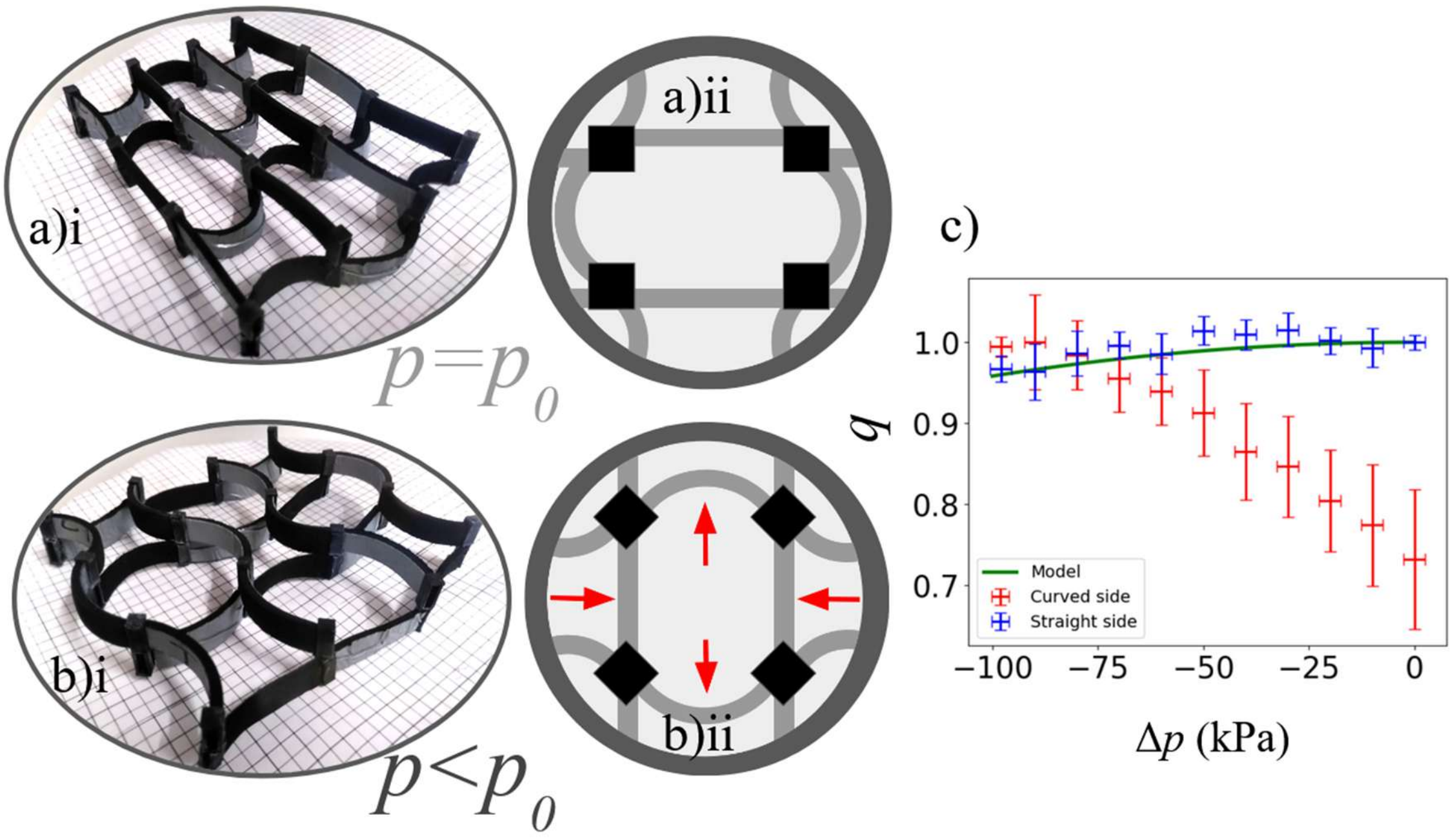

**Figure 6.** A metamaterial mesh, at pressure relative to atmosphere $\Delta p = 0$ (a) and $\Delta p = -98$ kPa (b), for both the pre-stressed (curved at $\Delta p = 0$) and straight sides. Overall shape change is visible (i), and diagrams illustrate bending at the level of individual cells (ii) – red arrows indicate bending direction at the middle of the actuators. c) Illustrates the difference in performance and uncertainty between the straight (negative compressibility) and curved (positive compressibility) sides, compared to the model in Equation 9 (for linear contraction factor $q$).

Interestingly, the resulting structure exhibited *negative compressibility*, as its length and width decreased as $p$ was reduced – in most materials, including the bare EPDM foam from which the actuators have been constructed, the reverse is true. While negative compressibility can be achieved structurally [23], using foam bi-layers for the purpose allows a greater material response to be achieved for a given $\Delta p$, and a greater potential for tuning of properties, in addition to their intrinsic advantages in size, weight and complexity.

The shape-changing properties of the meshes can be reconfigured not just by changing $l$, but also by replacing either the rows, columns or both with actuators which are curved at $\Delta p = 0$, rather than straight. This has the effect of switching the compressibility from positive (expanding with decreasing pressure) to negative (vice-versa) in one or two dimensions.

Pre-stressing actuators in their curved state can be performed by attaching the film to a long, thin strip of foam which is coiled under stress. Individual actuators may then be cut from this coil. This method of manufacture does not easily allow for the use of a laser cutter or a large, flat surface on which to precisely cut actuators. Thus, manufacturing precision is reduced, leading to greater standard deviations in $q$ for meshes made with pre-curved actuators. The radii at both atmospheric and vacuum pressures are also more difficult to control, as these are dependent on how the coil is formed, and may vary across the length of the strip. However, significant, consistent shape change was observed across repeats (Figure 6). Uncertainties, taken as the standard deviation across four repeats, were much greater for the pre-stressed sides due to the sources of error described. Whilst the straight sides continued to act in accordance with the model in Equation 9, while the curved sides exhibited a much higher rate of shape change. The mechanism by which this occurs is a potential avenue for new research.

Thus, these actuators can be used to create a fully customisable mesh – that is, one in which $q$ can be varied not only in magnitude (by changing $l_{arc}$), but also in sign (by pre-stressing), and both x- and y-dimensions can be varied independently. By altering actuator properties across the mesh, further control could be achieved. Metamaterials based on these actuators therefore allow pressure-based deformation to be precisely defined. This concept has great potential in the design of under sea pipes, submersibles, and other structures, as well as devices in other extreme pressure environments. Constructing actuators from metal foams [24] would allow deployment in higher-pressure under-sea regions, where EPDM actuators would not survive.

Furthermore, in all cases the nodes joining the actuators rotate as the mesh changes shape - this occurs in the opposite direction to a node's nearest neighbours. This has numerous potential applications – if each node were connected to an electromagnetically resonant element, the properties of an electromagnetic array could respond to atmospheric changes. Rudimentary screens have also been constructed using this principle, but actuated by other means, by placing light filters on the nodes [25] – similar approach utilising the ambient pressure response has much potential here also.

## 3 Lifting force

In addition to the change in shape, the force these actuators can supply is also highly relevant, especially if they are to be used to deploy larger structures. To measure the lifting force $F_\uparrow$ applied by an actuator at a given pressure, masses were hung from one end, with the other end fixed in place. The pressure within the chamber was then decreased, causing $F_\uparrow$ to build until the end of the actuator was raised above the ground. At the point at which the actuator end begins to move upwards, $F_\uparrow$ is exactly equal to the downward force $F_\downarrow$ due to the masses, equal to the mass of these multiplied by the gravitational acceleration $g = 9.81$ ms$^{-2}$. At the pressure this occurs, the lifting force can then be obtained by multiplying the total mass by $g$.

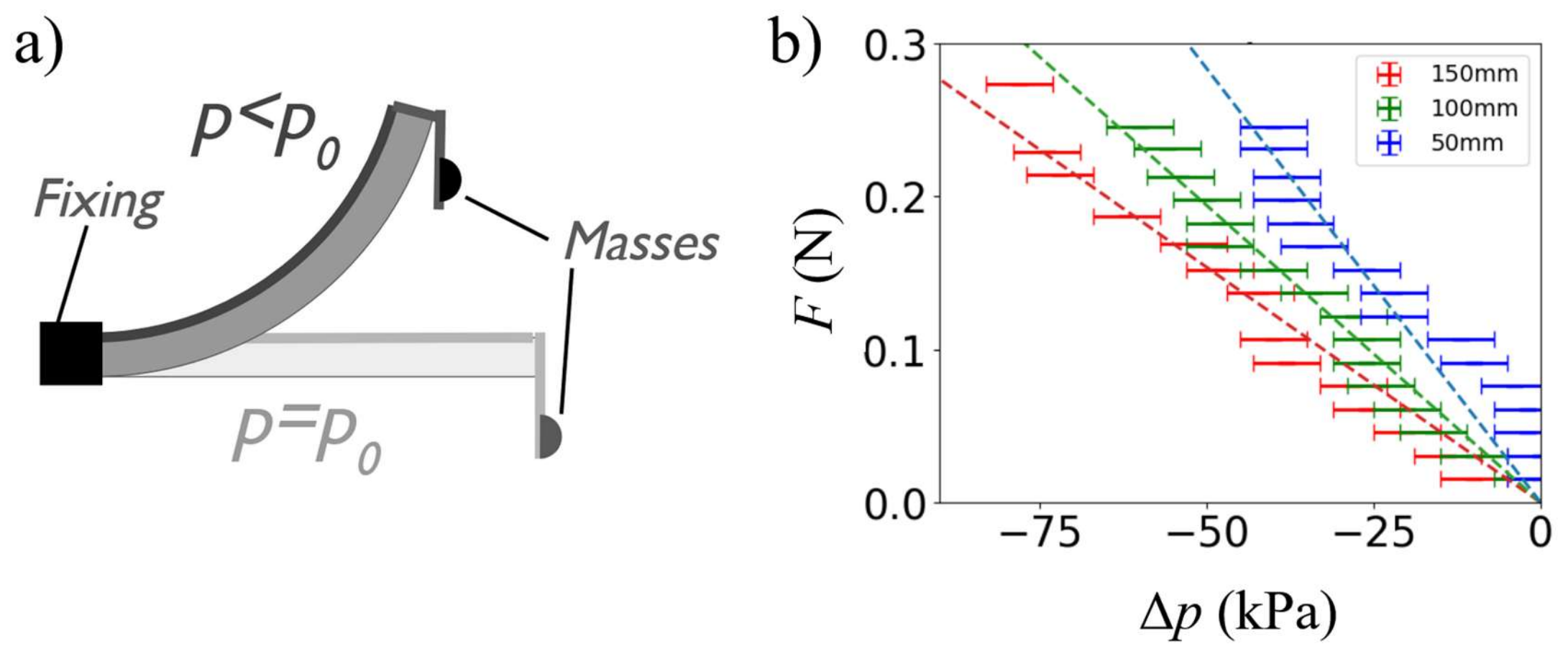

**Figure 7.** Setup for the measurement of force (a), and the force-pressure relation (b).

Unlike in the measurements of *r*, a strong dimensional dependence was observed (Figure 7), and the force each actuator supplied was found to vary linearly with $\Delta p$. The observation that shorter actuators are able to supply more force than longer ones is most likely due to the mass inducing a bending moment opposing the pressure-controlled actuator bending, related to the distance *d* between the fixed end and the centre of mass.

Uncertainties were taken as the uncertainties in measurement of mass and pressure – the latter was particularly significant, as this involved judging the exact point at which an actuator leaves the ground.

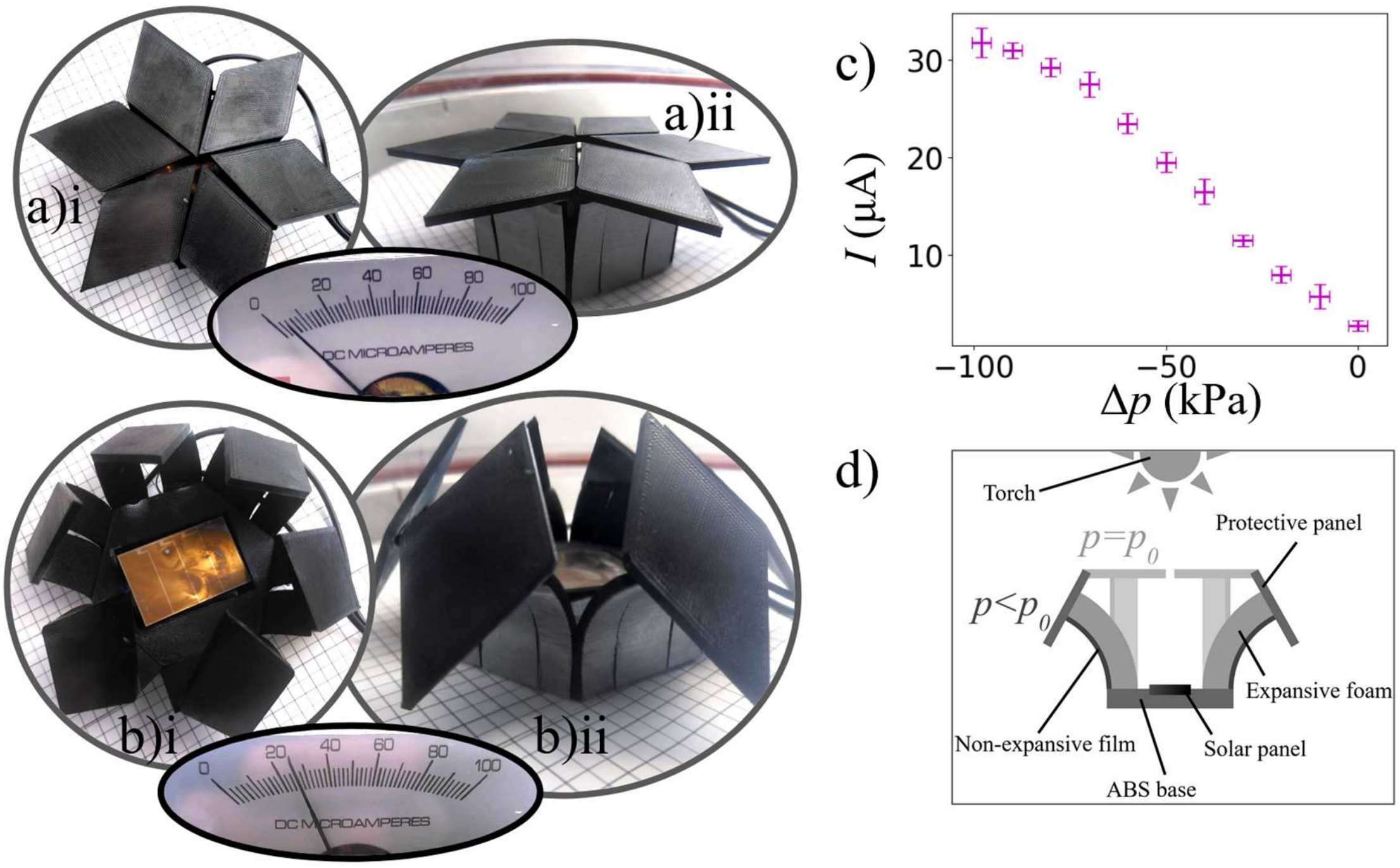

**Figure 8.** Top (i) and side (ii) views of the solar shield, with panel in situ, at atmospheric (a) and vacuum (b) pressures. (c) shows how the current $I$ through the panel varies as a function of pressure relative to the atmosphere $\Delta p$. The current measured through the panel increases from 4 μA to 26 μA as the shield opens. (d) shows a cross-section diagram of the device.

Using information gained around the relationship between actuator length and lifting force, a demonstrator device, which could open to reveal a solar panel, was designed and constructed. Twelve actuators, of dimensions $17.5 \times 30.0$ mm, were built and affixed upright to the six sides of a hexagonal 3D-printed ABS base, incorporating a small solar panel (of original design – see Figure 8d). To the ends of each pair of actuators, ABS 'petals' were attached, positioned to cover the panel completely. The mechanism was designed so that, at vacuum pressure, the actuators bent outwards (to $r = 31$ mm), opening the petals and completely exposing the panel to the light of an 800 lumen LED torch suspended 300 mm above the vacuum chamber.

The limits on device size were set by the joint needs to both accommodate a small commercial solar panel, and open fully within the confines of the vacuum chamber. One

Panasonic Amorphous Solar Cell was employed, measuring $41.6 \times 26.3 \times 1.3$ mm, which had a nominal open circuit voltage of 2.4 V. This was verified through experiment. $l$ was carefully chosen not only to ensure that these size demands were met, but also to make sure that, at $\Delta p = -98$ kPa, the petals would not block the panel (a calculation dependent on $r$, $l$ and the petal dimensions). Once this had been set, $w$ and dimensions of the rest of the structure were designed, ensuring that the mass of the petals ($4.0 \pm 0.5$ g) would not apply a $F_{\downarrow}$ exceeding $F_{\uparrow}$ (using data from Figure 7b).

Figure 8 shows that the foam actuators were able to successfully open and close the shield, producing a noticeable increase in current upon opening (Figure 8a-b) over multiple repeats. It should be noted that the ammeter reading did not start from zero as there were small gaps in the shield at panel joins. The current through the panel increased linearly with $\Delta p$, before tailing off at the end as the panel was completely exposed (Figure 8c). The demonstrator therefore also functions well as a simple pressure sensor, and has the potential to deploy or uncover various pieces of scientific equipment. Using the knowledge gained from previous sections, it is now possible to build a device with precisely tuned actuation force $F_{\uparrow}$ and linear expansivity $q$ for a variety of applications.

## 4 Conclusions

Shape-morphing bi-layer actuators controlled by ambient pressure have been constructed from EPDM rubber foam and a non-expansive film. These have been accurately characterised in terms of their expansivity, bending radius and linear contraction factor, both as a function of the ambient pressure and the actuator length. Configurable actuator metamaterial meshes, which can contract or expand in one or two dimensions, and a deployable device, were constructed and tested, both performing successfully and in line with theory.

Both ideas could be developed further, and tested in different or less controlled environments. For example, the mesh could be taken underwater to test the effects of positive pressure, and a derivative of the solar panel device could be tested in a zero-gravity environment, such as space. A mathematical model could also be developed for the lifting force, and how this varies with actuator dimensions. Other foam types could also be investigated. Ultimately, this technology may be employed to deploy structures and equipment in a variety of extreme environments, develop sensors for numerous stimuli, and allow large structures to morph in response to pressure – creating machines which, like plants, can optimise their shape to ambient conditions.

**Acknowledgements**

Alex Powell acknowledges support from a Royal Academy of Engineering Research Fellowship grant.

Table of references